\documentclass[9pt,twocolumn,twoside]{opticajnl}
\journal{ol}
\setboolean{shortarticle}{true}
\usepackage{lineno}

\title{Non-resonant Bragg scattering four-wave-mixing at near visible wavelengths in low-confinement silicon nitride waveguides}
\author[1]{Nicholas Jaber}
\author[1]{Scott Madaras}
\author[1]{Andrew Starbuck}
\author[1]{Andrew Pomerene}
\author[1]{Christina Dallo}
\author[1]{Douglas C. Trotter}
\author[1]{Michael Gehl}
\author[1]{Nils Otterstrom}
\affil[1]{Photonic and Phononic Microsystems, Sandia National Laboratories, Albuquerque, New Mexico 87185 USA}
\affil[*]{Corresponding author: njjaber@sandia.gov}

\begin{abstract}
Quantum state coherent frequency conversion processes—such as Bragg scattering four wave mixing (BSFWM)—hold promise as a flexible technique for networking heterogeneous and distant quantum systems. In this letter, we demonstrate BSFWM within an extended (1.2-m) low-confinement silicon nitride waveguide and show that this system has the potential for near unity quantum coherent frequency conversion in visible and near-visible wavelength ranges. Using sensitive heterodyne laser spectroscopy at low optical powers, we characterize the Kerr coefficient ($\sim$1.55 $\rm W^{-1}m^{-1}$) and linear propagation loss ($\sim$0.0175 dB/cm) of this non-resonant waveguide system, revealing a record-high nonlinear figure of merit (NFM = $\frac{\gamma}{\alpha} \approx$ 3.85 $\rm W^{-1}$) for BSFWM of near visible light in non-resonant silicon nitride waveguides. We demonstrate how, at high yet achievable on-chip optical powers, this NFM would yield a comparatively large frequency conversion efficiency, opening the door to near-unity flexible frequency conversion without cavity enhancement and resulting bandwidth constraints.
\end{abstract}

\setboolean{displaycopyright}{true}

\begin{document}
\maketitle

\section{Introduction}
\par The ability to interface heterogeneous quantum systems though quantum transduction operations has the potential to address key scaling challenges within quantum information platforms \cite{brown2016co, monroe2014large, monroe2013scaling,kumar1990quantum,ates2012two}. Particularly, quantum state coherent frequency conversion processes, such as BSFWM, could enable networking between disparate atomic, defect, and other photonically addressable quantum systems \cite{tanzilli2005photonic,monroe2014large,kumar1990quantum,sangouard2009quantum}. BSFWM is a nonlinear optical process that can be leveraged in systems with a high $\chi^{(3)}$ susceptibility (characterized by a system’s Kerr coefficient) to induce a polarization field coupling between signal and idler wavelengths for quantum transduction operations \cite{boyd2020nonlinear, mcguinness2010quantum}. Resonant BSFWM has been used to demonstrate high conversion efficiencies in compact structures at near IR-wavelengths, but resonance constraints in these systems naturally limit the availability of addressable frequencies \cite{li2016efficient,agha2012low,singh2019quantum}. 

\par In this letter, we demonstrate BSFWM within a non-resonant silicon nitride waveguide to convert photons with wavelengths of $\sim$784.55 nm to $\sim$830 nm. These measurements show a record-high nonlinear figure of merit (NFM = $\frac{\gamma}{\alpha} \approx$ 3.85 $\rm W^{-1}$) for BSFWM of near visible light in non-resonant silicon nitride waveguides. A system's NFM is exponentially related to the system’s suitability for producing frequency conversion operations at near unity system efficiency. This platform is CMOS fabricated and allows for simple integration with other CMOS compatible quantum technologies \cite{graham2014system,ivory2021integrated,mcguinness2022integrated}. Additionally, this non-resonant system permits conversion of photons between nearly arbitrary phase-matched frequencies given flexible pump wavelengths.

\section{Phase Matching Conditions for Bragg Scattering Four Wave Mixing}
\par
\begin{figure}[ht]
\centering
{\includegraphics[width=\linewidth]{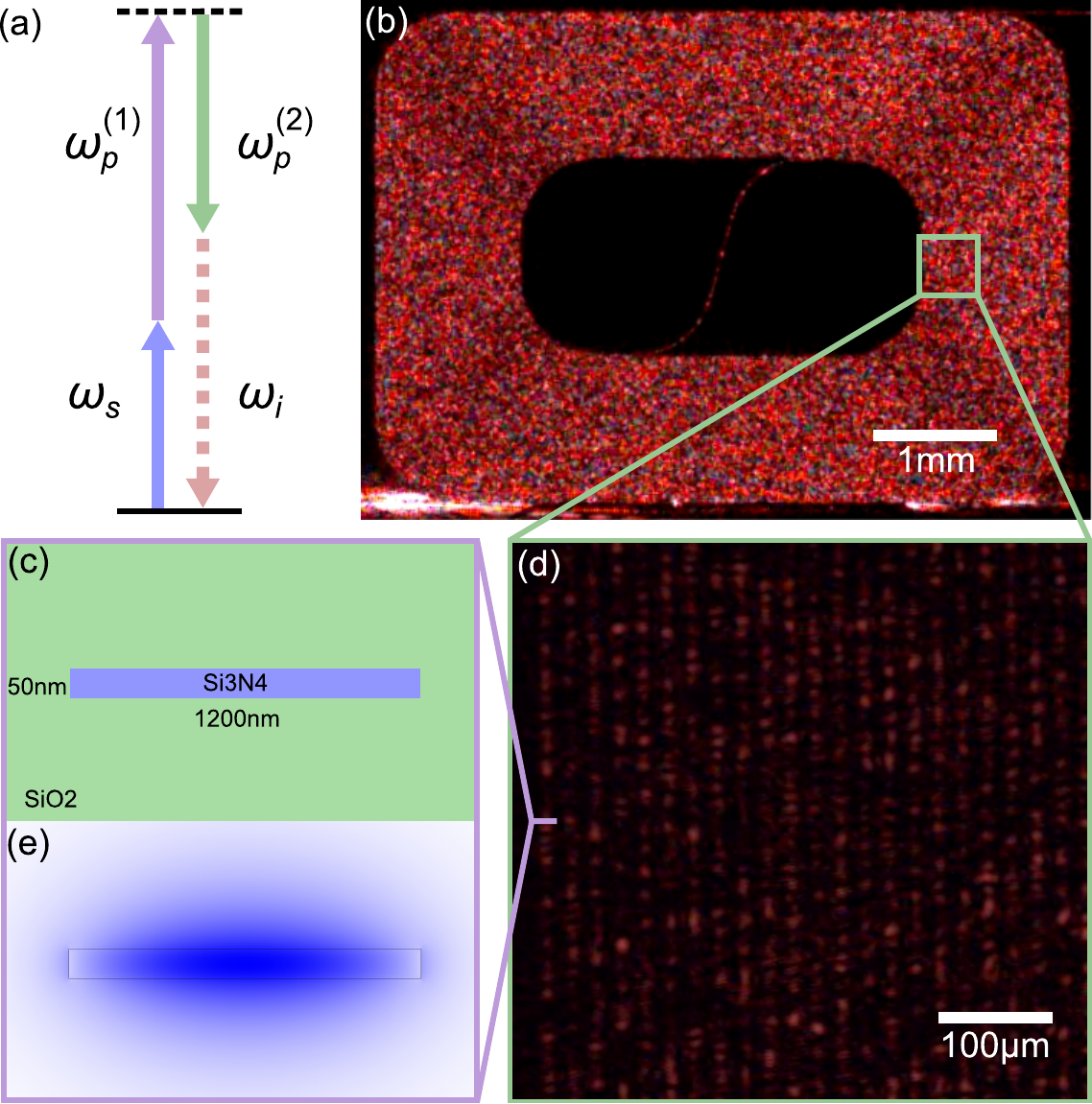}}
\caption{(a) Energy conservation conditions for the signal (blue), pump (purple and green), and idler photons (red). The dashed arrow for the idler photon indicates that it is generated through the BSFWM process. (b) Top down imaging of scattered light from the silicon nitride spiral waveguide. (c) Cross section diagram of the $\rm Si_3 N_4$ waveguide. (d) Micrograph of scattered light from spiral waveguide segments. (e) COMSOL finite element method (FEM) simulation of the fundamental TE-like mode at 830 nm in the waveguide.}
\label{fig:conversion}
\end{figure}

 BSFWM is a quantum state coherent frequency conversion process governed by energy conservation ($\omega_s + \omega_{p}^{(1)} = \omega_i + \omega_{p}^{(2)}$, see Fig. \ref{fig:conversion}a) and phase matching conditions ($k_s + k_{p}^{(1)} = k_i + k_{p}^{(2)}$), where $\omega_s$, $\omega_i$, $\omega_{p}^{(i)}$, and $k_s$, $k_i$, $k_{p}^{(i)}$ are the frequencies and wavenumbers of the signal, idler, and pump fields, respectively \cite{boyd2020nonlinear}. In BSFWM, pump fields ($\omega_{p}^{(1)}$, $\omega_{p}^{(2)}$) induce a coupling between signal ($\omega_s$) and idler frequencies ($\omega_i$). Formally this may be understood as process in which input frequencies ($\omega_s$, $\omega_{p}^{(1)}$) excite a short lived virtual state that rapidly decays into output photons ($\omega_i$, $\omega_{p}^{(2)}$) \cite{boyd2020nonlinear}.  

\par The system efficiency (F) and conversion efficiency ($\eta$) of BSFWM processes are dependent on the system’s pump powers ($P_1$ and $P_2$), Kerr coefficient ($\gamma$), phase mismatch ($\Delta k = k_s + k_{p}^{(1)} - k_i - k_{p}^{(2)}$), linear propagation loss ($\alpha$) and interaction length (z). This relationship is shown in Eq. \ref{eq:full_eq}, where $\kappa = 2 \gamma \sqrt{P_1 P_2}$, and $\delta \beta = \frac{1}{2}\Delta k + \gamma (P_1 - P_2)$ \cite{mcguinness2010quantum,otterstrom2021nonreciprocal,boyd2020nonlinear}. 

\begin{equation}
\begin{aligned}
     F &= \eta e^{-\alpha z} =  \frac{ \kappa^2 \sin^2[z \sqrt{\kappa^2 + \delta \beta^2}]}{\kappa^2 + \delta \beta^2} e^{-\alpha z}
\end{aligned}
\label{eq:full_eq}
\end{equation}

\section{Methods \& Results}
\par We fabricated a 1.2-m spiral waveguide using low pressure chemical vapor deposition (LPCVD) $\rm Si_3 N_4$, photolithography and inductively coupled plasma (ICP) etching in Sandia National Laboratories’ Silicon Fab at the Microsystems Engineering, Science and Applications (MESA) Complex. The silicon nitride waveguide has a thickness and width of 50 nm and 1200 nm, respectively, and is embedded within a high density plasma (HDP) $\rm SiO_2$ cladding. Figure \ref{fig:conversion}c,e show the waveguide cross-section and the corresponding fundamental TE-like mode supported in the structure. A top-down image/micrograph of light coupled into the spiral device is shown in Fig, \ref{fig:conversion}b,d. 

\par
\begin{figure}[t]
\centering
{\includegraphics[width=\linewidth]{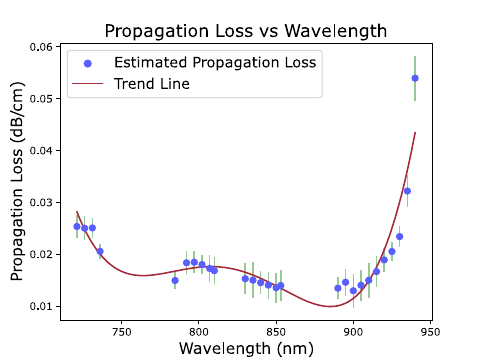}}
\caption{Measured propagation loss in units of dB/cm using a top-down micrograph technique. Loss was measured at experimentally available wavelengths at 5 nm increments between 721 nm and 940 nm. The inscribed trend line is a 4th-order polynomial fit and serves as a guide to the eye.}
\label{fig:loss}
\end{figure}

 We measure the linear propagation loss of our device using top-down imaging of light scattered from the spiral together with image processing methods described by Okamura et al. \cite{okamura1983measuring,mcguinness2022integrated}. As shown in Fig. \ref{fig:loss}, we consistently observe a loss of $\sim$0.0175 dB/cm for relevant wavelengths between 721 nm and 940 nm. We attribute the sharp increase in estimated propagation loss for wavelengths above 900 nm to bend induced losses, due to the low confinement of our waveguide at these wavelengths. These ultra-low loss waveguides allow for long nonlinear optical interaction lengths, holding promise for high system efficiencies using BSFWM of near visible light in a non-resonant system.

\par
\begin{figure}[tp!]
\centering
{\includegraphics[width=\linewidth]{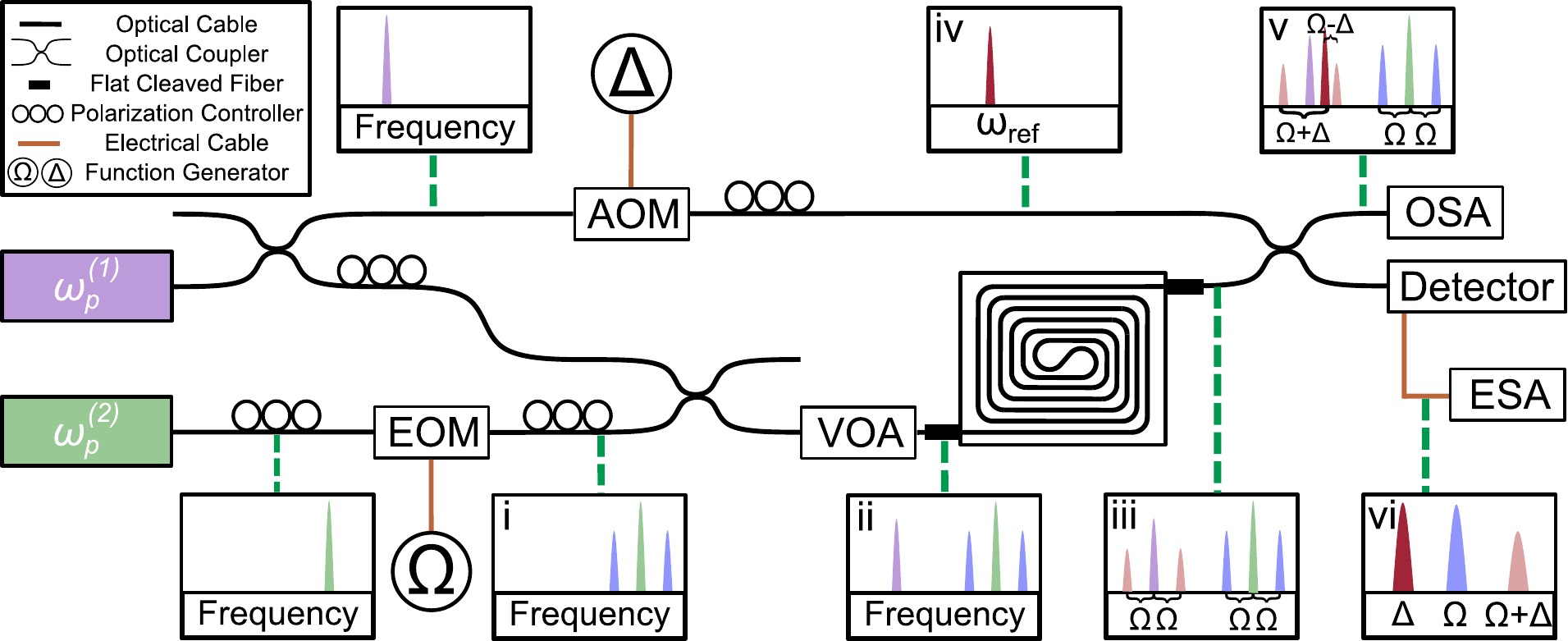}}
\caption{Laser heterodyne spectroscopy setup used to characterize the Kerr coefficient of the silicon nitride waveguide. Pump frequencies ($\omega_{p}^{(1)}$, $\omega_{p}^{(2)}$) originate from two distinct sources.  Light at $\omega_{p}^{(2)}$ is modulated by an EOM with an RF drive frequency of $\Omega$ to generate the signal field at $\omega_s$ = $\omega_{p}^{(2)} \pm \Omega$. The signal field is coupled with those of the pumps and passed through a VOA to control the on-chip optical power. These attenuated frequencies are edge coupled into the waveguide using a flat cleaved fiber. Throughout our device BSFWM converts $\omega_s$ and $\omega_{p}^{(1)}$ to $\omega_i$ and $\omega_{p}^{(2)}$. The output of this waveguide is edge coupled into another flat cleaved fiber, and coupled with a reference optical local oscillator (LO) derived from AOM-modulated light ($\omega_{p}^{(1)}$ + $\Delta$). Finally, we use a high-speed detector to mix relevant optical fields ($\omega_i$ and $\omega_{p}^{(1)}$ + $\Delta$, $\omega_{p}^{(2)}$ and $\omega_{p}^{(2)} \pm \Omega$), producing RF beat notes at frequencies $\Delta + \Omega$ and $\Omega$. The pump, signal, and idler light is also measured using an OSA to quantify the relative on-chip power of frequencies near $\omega_{p}^{(1)}$, and $\omega_{p}^{(2)}$.}
\label{fig:layout}
\end{figure}

 To characterize the waveguide's nonlinear optical Kerr coefficient, we use the laser heterodyne spectroscopy setup shown in Fig. \ref{fig:layout}. Signal light ($\omega_{s}$) is generated through intensity modulation of pump light ($\omega_{p}^{(2)}$) such that $\omega_{s} = \omega_{p}^{(2)} \pm \Omega$ (see Fig. \ref{fig:layout}i), where $\Omega$ is the RF frequency of the electro-optic modulation (EOM).  Pump and signal light ($\omega_{p}^{(1)}$, $\omega_{p}^{(2)}$, $\omega_{s}$) are then combined and passed through a variable optical attenuator (VOA) to control optical power levels (Fig. \ref{fig:layout}ii), and are subsequently coupled on-chip to produce idler light ($\omega_i$) through BSFWM (Fig. \ref{fig:layout}iii).  Following the interaction within the integrated photonic spiral, the light is coupled off chip and combined with a reference optical local oscillator (LO) (Fig. \ref{fig:layout}v), which is generated through acousto-optic modulation (AOM) of pump light ($\omega_{p}^{(1)}$) (see $\omega_{ref}$ in Fig. \ref{fig:layout}iv).  The combined optical fields produce distinct microwave signals (Fig. \ref{fig:layout}vi) through high-speed RF detection of the optical beat notes, which we collect using an electrical spectrum analyzer (ESA). 

\par In our experiment, we set pump wavelengths to $\lambda _{p}^{(1)} = 830$ nm and $\lambda _{p}^{(2)} = 784.55$ nm ($\lambda_{p} = 2\pi c / \omega_{p}$), $\Omega$ = 60 MHz, and $\Delta$ = 117 MHz. Using the same finite element method (FEM) simulation in Fig. \ref{fig:conversion}e, we calculate that for these parameters $\Delta k \approx 0.0507$ $\rm m^{-1}$, which is well within the phase matching limit ($\Delta kz \approx 1$). Within this limit, our 1.2-m spiral waveguide can support BSFWM with signal-pump detuning ($\Omega$) up to 980 MHz.

\par The conversion efficiency is extracted by comparing the heterodyne electrical signals and relative powers of the optical fields.  In particular, we measure the generated idler power by examining the RF power at frequency $\Omega + \Delta$ ($P_{\Delta + \Omega}^{\rm RF}$), produced by mixing the idler frequency ($\omega_{i}$) and reference optical LO frequency $\omega_{ref}$ (Fig. \ref{fig:layout}vi). In order to normalize this data, we measure the beat note power ($P_{\Omega}^{\rm RF}$) between pump ($\omega_{p}^{(2)}$) and signal light, appearing at an RF frequency of $\Omega$ (Fig. \ref{fig:layout}vi). We normalize/reference these beat notes with the relative optical powers of the pump ($P_{2}$) and the reference LO ($P_{ref}$), as measured by an optical spectrum analyzer (OSA) (Fig. \ref{fig:layout}v). Putting this together, we compute the conversion efficiency as $\eta = P_{\Delta + \Omega}^{\rm RF} P_{2} / (P_{\Omega}^{\rm RF} P_{ref})$. Using this approach in conjunction with the VOA, we measure the BSFWM conversion efficiency as a function of on-chip pump power ($P_{1}$, $P_{2}$) to characterize the waveguide's Kerr coefficient.

\par
\begin{figure}[t]
\centering
{\includegraphics[width=\linewidth]{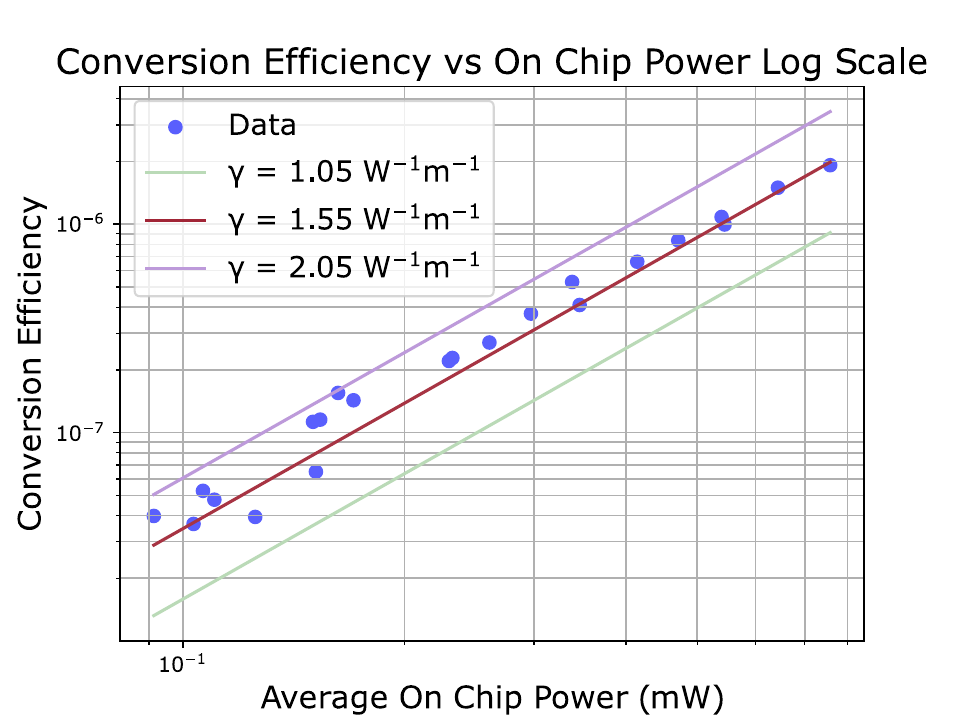}}
\caption{Measured relationship between conversion efficiency and on-chip pump power, demonstrating good agreement between the experimental data and BSFWM theory with a Kerr coefficient of $\gamma$ = 1.55 $\rm W^{-1}m^{-1}$.}
\label{fig:kerr}
\end{figure}

 Figure \ref{fig:kerr} demonstrates good agreement between the experimental data and BSFWM theory with a Kerr coefficient of $\gamma$ = 1.55 $\rm W^{-1}m^{-1}$. Combining this Kerr coefficient with the demonstrated ultra-low propagation loss ($1.75$ $\rm dB/m$, $\alpha = 0.403$ $\rm m^{-1}$), this device achieves a record-high NFM ($\frac{\gamma}{\alpha} \approx$ 1.55 $\rm W^{-1}m^{-1}$/ 0.403 $\rm m^{-1}$ = 3.85 $\rm W^{-1}$) for BSFWM of near visible light in non-resonant silicon nitride waveguides. In conjunction with the calculated effective area from simulation data (Fig. \ref{fig:conversion}e), we estimate our LPCVD $\rm Si_3 N_4$ Kerr nonlinearity as $\rm n_2 = 4.16 \times 10^{-19} m^2 W^{-1}$. Within the uncertainty of our on-chip pump power estimation, this value is consistent with published work on similar $\rm Si_3 N_4$ films \cite{gao2022probing,kruckel2017optical}. 

\section{Discussion}
\par
\begin{figure}[tp!]
\centering
{\includegraphics[width=\linewidth]{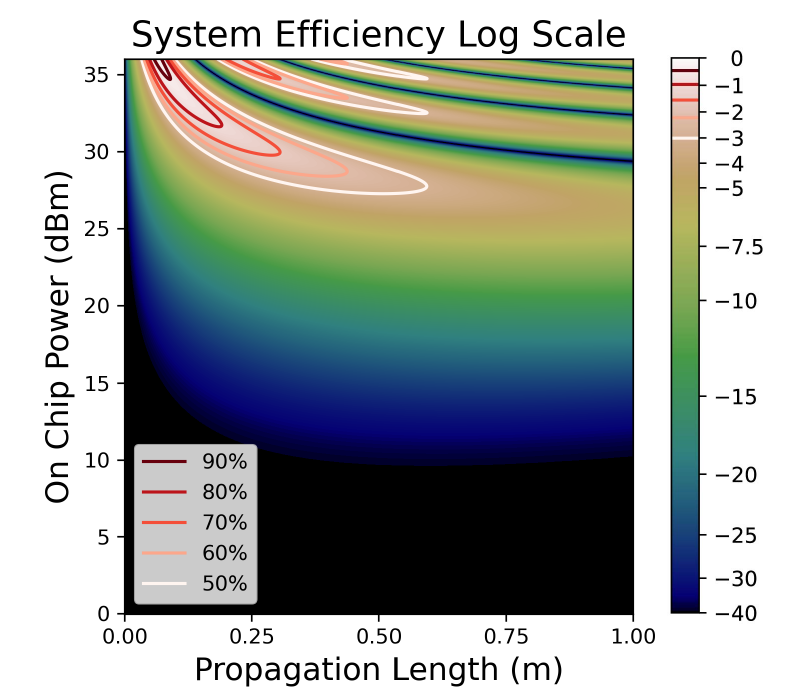}}
\caption{Theoretical effect of interaction length and optical power on system efficiency. The contour lines circumscribe  the parameter space capable of high system efficiencies (>50\%).}
\label{fig:computational}
\end{figure}

 The combination of strong Kerr nonlinearity with ultra-low propagation loss holds promise for high system efficiency quantum frequency conversion. In Fig. \ref{fig:computational}, we estimate the system efficiency (Eq. \ref{eq:full_eq}) under various powers and propagation lengths based on the characterization of our device at low pump power in Fig. \ref{fig:layout}. We identify points of high system efficiency (>70\%), such as 1 W (30 dBm) of on-chip power and 24 cm of propagation length. If the application is amenable to time gating, this high system efficiency is possible at low average pump powers using a pulsed pump scheme. For instance, a 1\% duty cycle with a peak power of 1 W corresponds to an average pump power of 10 mW, well within the power handling capabilities of this waveguide. This marks a clear route to viability as a near unity flexible frequency conversion platform for near visible light.

\par Building on this work, we can adapt this platform to convert photons between telecom and visible frequencies, opening the door to high efficiency quantum networking with distant and disparate quantum systems. To expand the range of viable phase matching conditions, intermodal scattering processes---in which pump and signal/idler light are in distinct spatial modes of the waveguides---can be used to enhance the separation between pump and idler frequencies (from $\sim$1 GHz to several THz) \cite{otterstrom2021nonreciprocal,lacava2019intermodal,xiao2014theory}. Using telecom wavelengths to transmit through existing telecommunications infrastructure and a standard quantum networking protocol, this form of quantum frequency conversion may enable all to all quantum networking of photonically addressed quantum systems \cite{tanzilli2005photonic,monroe2014large,kumar1990quantum,sangouard2009quantum,monroe2013scaling,ates2012two,agha2013chip}.

\section {Funding}
\par This material is based upon work supported by the Laboratory Directed Research and Development program at Sandia National Laboratories. This article has been authored by an employee of National Technology \& Engineering Solutions of Sandia, LLC under Contract No. DE-NA0003525 with the U.S. Department of Energy (DOE). The employee owns all right, title and interest in and to the article and is solely responsible for its contents. The United States Government retains and the publisher, by accepting the article for publication, acknowledges that the United States Government retains a non-exclusive, paid-up, irrevocable, world-wide license to publish or reproduce the published form of this article or allow others to do so, for United States Government purposes. The DOE will provide public access to these results of federally sponsored research in accordance with the DOE Public Access Plan https://www.energy.gov/downloads/doe-public-access-plan.

\bibliography{hifi_bib}
\end{document}